\documentclass[preprint,nofootinbib,12pt,floatfix,aps,prd,superscriptaddress]{revtex4-1}
\usepackage{amsmath,amssymb,amsfonts,amstext,amsbsy,amsopn}
\usepackage{graphicx, color, xcolor, slashed, esint}
\usepackage[dvips]{epsfig}
\usepackage{float, units, textcomp, hyperref}
\usepackage[utf8]{inputenc}
\usepackage{subcaption}
\usepackage[normalem]{ulem}
\usepackage{orcidlink}
\newcommand{\beq}{\begin{equation}}
\newcommand{\eeq}{\end{equation}}
\newcommand{\bea}{\begin{eqnarray}}
\newcommand{\eea}{\end{eqnarray}}

\begin{document}

\title{Regular Black Strings and $(2+1)$ Black Hole in Unimodular Gravity Supported by Maxwell Fields}
\author{G. Alencar
\orcidlink{0000-0002-3020-4501}
}
\email{geova@fisica.ufc.br}
\affiliation{Department of Physics, Universidade Federal do Cear\'a (UFC), Campus do Pici, Fortaleza - CE, C.P. 6030, 60455-760 - Brazil}
\author{V. H. U. Borralho
\orcidlink{0009-0009-0085-6152}
}
\email{victorborralho@fisica.ufc.br} 
\affiliation{Department of Physics, Universidade Federal do Cear\'a (UFC), Campus do Pici, Fortaleza - CE, C.P. 6030, 60455-760 - Brazil}

\begin{abstract}
In this work, we obtain  regular solutions with a dynamical cosmological function in unimodular
gravity.  This alternative theory to General Relativity imposes an additional condition on the spacetime volume element by fixing the metric volume density to a prescribed non-dynamical quantity, which can be represented by a constant in unimodular coordinates. In this procedure, the cosmological constant does not appear directly in the action, but rather as an integration constant of the field equations. 
By using the non-conservation of the energy-momentum tensor, we show that the integration constant becomes a function $\Lambda(x)=\Lambda_0+\delta\Lambda(x)$, where $\Lambda_0$ is the vacuum cosmological constant and $\delta\Lambda(r)$ is a matter-induced cosmological contribution determined by the geometry and the electromagnetic sector.
From the definition of the geometric function $H(r)$, we verify the validity of Maxwell electrodynamics as a source for the solutions. We further show that linear combinations of the mass functions associated with the individual solutions can also be consistently supported by the Maxwell field, a property that is not generally shared by nonlinear electrodynamics. This result allows us to construct a broader family of regular solutions and investigate their physical properties. In particular, we analyze the energy conditions, curvature invariants, and thermodynamic properties of the generated solutions.
\end{abstract}
%
%
\maketitle
\section{Introduction}
In a recent work, Alencar \cite{Alencar:2026EEP} suggested that the Einstein Equivalence Principle leads to a fixed spacetime volume density, which in unimodular coordinates can be expressed as the condition $\det(g_{\mu\nu})=const.$, making it compatible with unimodular gravity and indicating that the cosmological constant may emerge as a consequence of this fundamental postulate.
Essentially, the resulting solutions coincide with those of general relativity; however, the absence of the trace equation leads to a different differential structure, where an integration constant can be identified with the cosmological term. 

While unimodular gravity is classically equivalent to general relativity under the assumption of energy-momentum conservation, the two theories differ at the quantum level due to their distinct symmetry structures, leading to different treatments of vacuum energy and the cosmological constant \cite{Bufalo:2015wda,Jirousek:2023gzr,Alvarez:2023eqo,Smolin:2009ti}. In general relativity, the full invariance under diffeomorphisms implies the covariant conservation of the energy-momentum tensor. In the framework of unimodular gravity, the restriction to volume-preserving diffeomorphisms, associated with the condition $\det(g_{\mu\nu}) = const.$, reduces this symmetry. As a consequence, the conservation of the energy-momentum tensor is no longer automatically guaranteed and can be relaxed \cite{Astorga-Moreno:2019uin,Fabris:2021atr,Fabris:2021kxt,Daouda:2018kuo}.  In this context, unimodular gravity has been extensively explored in cosmological scenarios \cite{Gao:2014nia,Singh:2023jxu,Leon:2022kwn,Shaposhnikov:2008xb}, including models with extra spatial dimensions in $D=4+d$ dimensions \cite{Velasquez-Toribio:2026cdx}, as well as in higher-dimensional formulations \cite{Kluson:2023yzv,NISHINO2010382}. Moreover, unimodular gravity has been investigated in the context of quantum gravity \cite{Unruh:1988in,deBrito:2021pmw,Smolin:2010iq,Padilla:2014yea}, including the analysis of quantum gravitational effects in arbitrary spacetime dimensions \cite{Anero:2023xjb}.

The traceless gravitational equations have an important historical precedent in Einstein's 1919 investigation of gravitational and electromagnetic models of elementary particles \cite{Einstein:1919gv}. In the conservative sector considered there, the Bianchi identity, together with $\nabla^\mu T_{\mu\nu}=0$, forces the quantity arising from the trace equation to be a constant. The present construction does not modify this conclusion. Instead, we explore the non-conservative sector of unimodular gravity, for which $\nabla^\mu T_{\mu\nu}=\partial_\nu\Lambda$ allows the integration function $\Lambda$ to depend on spacetime. Thus, the difference from Einstein's construction lies in the conservation assumption rather than in a different treatment of the same conservative equations.

Black holes and black strings are fundamental solutions of General Relativity, but they are generically plagued by spacetime singularities, where curvature invariants diverge and the classical description breaks down. In $(2+1)$ dimensions, the BTZ black hole \cite{Banados:1992wn} provides a paradigmatic example, capturing many essential aspects of black hole physics in an asymptotically AdS setting; however, its charged extension also exhibits singular behavior. In (3+1) dimensions, black strings—cylindrically symmetric solutions with negative cosmological constant—were obtained by Lemos \cite{LEMOS} and can be viewed as extended counterparts of the BTZ geometry, sharing similar asymptotics but differing in their global structure and thermodynamic properties. These examples highlight that singularities are a robust feature across different dimensions and symmetries.

This problem has driven the development of regular geometries, in which the central singularity is replaced by a finite and well-behaved core \cite{Simpson:2018tsi,Bronnikov:2021uta,Bardeen1968,Hayward:2005gi}. The first example of a regular black hole was introduced by Bardeen \cite{Bardeen1968}, with its matter content later identified within the framework of nonlinear electrodynamics by Ayon-Beato and Garcia \cite{Ay_n_Beato_2000}. This approach has been successfully extended to other settings, including cylindrical symmetry, leading regular black string and $(2+1)$ solutions supported by nonlinear electrodynamics \cite{Cataldo:2000ns, Furtado:2022tnb, Lima:2022pvc,Lima:2023arg,Lima:2023jtl,Alencar:2026qeb,Alencar:2025yyl,Silva:2025fqj,Alencar:2024nxi}.

All the sources mentioned above are typically highly nonlinear, which often obscures the physical interpretation and complicates the analysis of the underlying dynamics. However, one of the present authors has recently shown that, within non-conservative unimodular gravity, it is possible to support regular black holes using standard Maxwell electrodynamics, with part of the complexity effectively absorbed by the dynamical cosmological function $\Lambda$ \cite{Alencar:2026EEP}. More recently, this mechanism has also been successfully applied to the construction of regular black hole and traversable wormhole geometries without the need for nonlinear electrodynamics \cite{3137491,Alencar:2026hjf}. In this work, we extend this idea to cylindrical and lower-dimensional settings, constructing regular black string and charged $(2+1)$ solutions sourced by Maxwell fields in UG. This paper is organized as follows: in In Sec. \ref{SEC2}, we construct the known black string and BTZ black hole solutions within the framework of unimodular gravity, presenting the main equations that characterize the formalism and establish the basis for the subsequent analysis. In Sec. \ref{sec3}, we couple nonlinear electrodynamics to gravity, and through the non-conservation of the energy-momentum tensor, we find that $\Lambda$ acts as an effective electromagnetic source. In Sec. \ref{sec4}, we apply this framework to regular black strings and regular black holes, highlighting the differences between the unimodular gravity approach and the nonlinear electrodynamics framework commonly employed to construct such regular solutions. In Sec. \ref{sec:physical_properties}, we analyze the physical properties of the family of solutions, including their regularity, thermodynamic behavior, and compatibility with the energy conditions. Finally, in Sec. \ref{sec5}, we summarize the results and present our final remarks.

\section{Unimodular Gravity framework} \label{SEC2}

In this framework, the unimodular condition can be implemented at the action level by introducing a Lagrange multiplier. In $n$ dimensions, we consider
\begin{equation}
S=\int d^n x\left[
\sqrt{-g}\left(R+\mathcal{L}_m\right)
-\lambda(x)\left(\sqrt{-g}-\epsilon_0(x)\right)
\right],
\end{equation}
where $\lambda(x)$ is a Lagrange multiplier and $\epsilon_0(x)$ is a prescribed non-dynamical volume density. Variation with respect to $\lambda(x)$ imposes the unimodular constraint
\begin{equation}
\sqrt{-g}=\epsilon_0(x).
\end{equation}
Variation with respect to the metric, followed by elimination of the Lagrange multiplier through the trace, yields the traceless field equations. Setting the gravitational coupling to unity, these are
\begin{equation}
R_{\mu\nu}-\frac{R}{n}g_{\mu\nu}
=
T_{\mu\nu}-\frac{1}{n}g_{\mu\nu}T ,
\end{equation}
with $T=T_{\mu}^{\mu}$.

It is important to clarify the meaning of the unimodular constraint in the coordinate systems used below. Since $\sqrt{-g}$ is a scalar density rather than a scalar, $\epsilon_0(x)$ can be represented by a numerical constant in unimodular coordinates, while in a general coordinate system it acquires the corresponding coordinate dependence. For the black-string metric in Eq.~\eqref{metric} and the BTZ metric in Eq.~\eqref{BTZmetric}, one has, respectively,
\begin{equation}
\sqrt{-g}=\frac{r^2}{\ell},
\qquad
\sqrt{-g}=r.
\end{equation}
In both cases, the volume density is independent of the dynamical metric function $f(r)$ and therefore represents the prescribed density $\epsilon_0(x)$ in these coordinates. Hence, the coordinate dependence of the metric determinant does not represent a violation of the unimodular constraint.

In the simplest case, we have $\nabla_{\mu}T^{\mu\nu}=0$ (General Relativity), and taking the divergence of the field equations leads to
\begin{equation} \label{trace}
    \partial_{\mu}[(n-2)R+2T]=0 \implies (n-2)R+2T=2n\Lambda.
\end{equation}
Thus, we obtain
\begin{equation}
    G_{\mu\nu} +\Lambda g_{\mu\nu}=T_{\mu\nu}.
\end{equation}
Imposing the non-conservation of the energy-momentum tensor, we now have
\begin{equation}\label{UG}
    G_{\mu\nu}+ \Lambda(x)g_{\mu\nu}=T_{\mu\nu}, \quad \nabla^{\mu}T_{\mu\nu}=\nabla_{\nu} \Lambda.
\end{equation}
Thus, the non-conservation of the energy-momentum tensor implies a position-dependent cosmological term, which can effectively absorb part of the matter contribution, allowing for simpler source configurations while maintaining the same gravitational dynamics. We stress that $\Lambda(x)$ is not an additional fundamental scalar field and no independent equation of motion is assigned to it. It arises as the integration function associated with the non-conservation relation $\nabla^\mu T_{\mu\nu}=\partial_\nu\Lambda$. It is useful to separate it as
\begin{equation}
\Lambda(x)=\Lambda_0+\delta\Lambda(x),
\end{equation}
where $\Lambda_0$ is the constant vacuum cosmological term, while $\delta\Lambda(x)$ is determined by the matter distribution and the geometry through the field equations. Thus, when $\Lambda(x)$ depends on the parameters characterizing the matter source, as in the solutions considered below, this dependence represents a matter-induced cosmological contribution rather than an independent vacuum component. Its physical significance is therefore not that of an additional fundamental field, but of the geometric response associated with the exchange between the non-conservative matter sector and the spacetime-dependent cosmological contribution.

Furthermore, considering the static and cylindrically symmetric nature of the black string metric, as well as the static and circularly symmetric nature of the BTZ black hole metric, all physical quantities must depend solely on the radial coordinate $r$. Consequently, the dynamical cosmological parameter reduces to a radial function, $\Lambda = \Lambda(r)$.


\subsection{Black string}
Now, we will obtain the vacuum black string solution using unimodular gravity. Consider a static metric with cylindrical symmetry \cite{LEMOS}
\begin{equation}\label{metric}
    ds^2=-f(r)dt^2+\frac{1}{f(r)}dr^2 + r^2 d\phi^2 + \frac{r^2}{\ell^2}dz^2,
\end{equation}
where $\ell$ is the fundamental length of the model. In vacuum, the right-hand side of the field equation is zero. Thus, Eq.\eqref{trace} for $n=4$ is given by,
\begin{equation}
    R^{\ \nu}_\mu - \delta_\mu ^\nu\frac{R}{4} =0.
\end{equation}

Computing the Ricci tensor components and the Ricci scalar, we have
\begin{equation}
    R^{\ t}_t=R^{\ r}_r=-\frac{f''(r)}{2}-\frac{f'(r)}{r},
\end{equation}
\begin{equation}
    R^{\ \phi}_\phi=R^{\ z}_z=-\frac{r f'(r)+f(r)}{r^2}.
\end{equation}
\begin{equation}
   R= -f''(r)-\frac{4 f'(r)}{r}-\frac{2 f(r)}{r^2}
\end{equation}

Since $R^{\ t}_t=R^{\ r}_r$ and $R^{\ \phi}_\phi=R^{\ z}_z$, (in vacuum) we fundamentally have only one independent equation, given by
\begin{align}
    \frac{f''(r)}{4}-\frac{f(r)}{2 r^2}=0,
\end{align}
whose solution is
\begin{equation}
    f(r)=\frac{C_1}{r}+C_2 r^2.
\end{equation}
The constant $C_1=-4 m\ell$ is related to the linear density of the black string, and the constant $C_2$ can be associated with the cosmological term: $C_2=-\Lambda_0/3$, with $\Lambda_0=-3/\ell^2$. Finally, we obtain the black string solution in vacuum using unimodular gravity
\begin{equation}
    f(r)=-\frac{4 m \ell}{r}+\frac{r^2}{\ell^2}.
\end{equation}

\subsection{Neutral BTZ Black Hole}
To obtain the BTZ black hole in unimodular gravity, we start from the $(2+1)$-dimensional metric \cite{Banados:1992wn}
\begin{equation}\label{BTZmetric}
    ds^2=-f(r)dt^2+\frac{1}{f(r)}dr^2+r^2 d\phi ^2.
\end{equation}
Taking $n=3$ in Eq. \eqref{trace}, we have
\begin{equation}
      R^{\ \nu}_\mu - \delta_\mu ^\nu\frac{R}{3} =0.
\end{equation}
Repeating the steps from the previous section, we obtain the following equations:
\begin{align}
  R^{\ t}_t&=R^{\ r}_r=-\frac{1}{4} f''(r) \\ 
   R^{\ \phi}_\phi&=\frac{r f''(r)-2 f'(r)}{4 r} \\
   R&=-f''(r)-\frac{2 f'(r)}{r}.
\end{align}
For the BTZ black hole, it is known that the Ricci scalar is a constant $C_0$. Thus, solving the equation above:
\begin{equation}
    f(r)=-\frac{C_0 r^2}{6}-\frac{C_1}{r}+C_2
\end{equation}
setting $C_1=0$, $C_2=-M$, and $C_0=6\Lambda_0$, where in $(2+1)$ dimensions $\Lambda_0=-1/\ell^2$. In this way, we obtain the BTZ metric using the framework of unimodular gravity:
\begin{equation}
    f(r)=-M+\frac{r^2}{\ell^2}.
\end{equation}
To obtain the charged solution, we can add the term $-2q^2\ln(r)$:
\begin{equation}
f(r)=-M+\frac{r^2}{\ell^2}-2q^2\ln(r),
\end{equation}
where the charge term introduces curvature singularity at the origin.  From the behavior of the Kretschmann scalar, we see that the origin corresponds to a physical curvature singularity,
\begin{equation}
   K= \frac{12}{\ell^4}-\frac{8q^2}{\ell^2 r^2}+\frac{12q^4}{r^4},
\end{equation}
which behaves as $K\sim r^{-4}$ as $r\to0$.
\section{non-conservative framework: nonlinear electrodynamics} \label{sec3}
In this section we will couple nonlinear electrodynamics to gravity, and thereby use a non-conservative framework so that we can obtain solutions with $\Lambda=\Lambda(x)$. For this, we add the general Lagrangian $L(F)$ to the Einstein-Hilbert action.
\begin{equation}
    S=\int d^4x \sqrt{-g}\Big( R + 2L(F) \Big),
\end{equation}
with $F=F_{\mu\nu}F^{\mu\nu}$. This framework is well-established in the literature, covering a vast array of physically significant NED models, notably those that yield regular or modified black object geometries.

Considering a generic nonlinear electromagnetic Lagrangian density $L(F)$, the associated energy-momentum tensor takes the form
\begin{equation} \label{tmunu}
    T_{\mu \nu} = g_{\mu\nu}\,\frac{L(F)}{2} - 2\,L_F\,F_{\mu \alpha} F_{\nu}^{\ \alpha}.
\end{equation}
with trace is $T=2L(F)-2FL_F$ (we have introduced the notation $L_F=dL/dF$).

By varying the action with respect to the gauge field $A^\mu$, we obtain the electromagnetic field equations. This procedure yields the generalized Maxwell equations governing nonlinear electrodynamics
\begin{equation}\label{mawxellgen}
    \nabla_\mu \left( L_F F^{\mu\nu} \right) = J^\nu.
\end{equation}

Thus, from the previous equation, we compute the divergence of the energy-momentum tensor:
\begin{equation}\label{lambda}
    \nabla^\mu T_{\mu\nu} = -2F_{\nu\alpha}J^\alpha \implies \partial_\nu \Lambda = -2 F_{\nu\alpha}J^\alpha,
\end{equation}
indicating that the matter-induced cosmological contribution is no longer a constant value, and can be interpreted as an effective electromagnetic source \cite{3137491}.

It is important to distinguish the conservation of the electromagnetic sector from that of the complete system. In the non-conservative sector considered here, the electromagnetic energy-momentum tensor is not conserved separately. Nevertheless, from the field equations and the Bianchi identity,
\begin{equation}
\nabla_\mu\left[T^{\mu\nu}-\Lambda(x)g^{\mu\nu}\right]=0.
\end{equation}
Therefore, the complete effective energy-momentum tensor is covariantly conserved, and the spacetime dependence of $\Lambda(x)$ accounts for the exchange between the electromagnetic and cosmological contributions. Note that $\nabla_\mu T^{\mu\nu}_{\rm EM}=-F^{\nu\lambda}J_\lambda$ is not a departure from standard electrodynamics: it is the usual Lorentz-force identity for a current-carrying electromagnetic field in curved spacetime. In ordinary (conservative) General Relativity with charged matter, this same non-conservation of $T^{\mu\nu}_{\rm EM}$ alone is compensated by the charged-matter stress tensor; here it is instead compensated by the dynamical vacuum sector $\Lambda(x)$, without invoking additional matter fields. The current $J^\mu$ appearing in Eq.~\eqref{mawxellgen} is, in this sense, the effective current associated with this non-conservative sector: it is the ordinary Gauss-law current, Eq.~\eqref{current}, required to sustain the electric-field profile $E(r)$ fixed by the regularity of the mass function, and is not an independently prescribed external source. This is also why, as noted below, purely magnetic configurations ($J^\mu=0$) force $\partial_r\Lambda=0$ and recover the standard conservative sector with constant $\Lambda_0$: the non-conservative behavior is turned on precisely by, and only by, the presence of a genuine electric current. If the additional source-free condition $J^\mu=0$ is imposed, Eq.~\eqref{lambda} gives $\partial_\nu\Lambda=0$, and the standard conservative sector with a constant cosmological term is recovered.

Physically, this relation describes a direct energy exchange between the nonlinear electromagnetic field and the spacetime vacuum, mediated by the effective current $J_\alpha$. In order to explicitly integrate the field equations and determine both the metric function $f(r)$ and the profile of $\Lambda(r)$, one must specify a particular model for the nonlinear electromagnetic Lagrangian $L(F)$.

In order to have $\Lambda=\Lambda(r)$, we must satisfy $\partial_r\Lambda(r) \neq 0$. In the magnetic configuration considered here, the only nonvanishing component of the field strength is $F_{23}\propto Q_m$. Therefore, from $\partial_\nu\Lambda=-2F_{\nu\alpha}J^\alpha$, the radial component gives $\partial_r\Lambda(r)=0$. Therefore, in this work, we will exclusively present electric-type solutions.

For a purely electric configuration, we take
\begin{equation}
F_{01}=-E(r),
\end{equation}
where $E(r)$ denotes the electric field. Consequently, the electromagnetic invariant is given by
\begin{equation}
F=F_{\mu\nu}F^{\mu\nu}=-2E^2(r).
\end{equation}
With this, we can generally write the field equations for a black string generated by nonlinear electrodynamics within the UG framework as:
\begin{align}\label{feBS}
   \Lambda(r)  -\Lambda_0 -\frac{4 m'(r) \ell}{r^2} &= \frac{1}{2}L(F)+ 2E^2(r) L_F , \\ 
 \Lambda(r)   -\Lambda_0 - \frac{2 m''(r) \ell}{r} &= \frac{1}{2}L(F) .
\end{align}

In Ref.\cite{3137491}, the geometric function $H(r)$ is defined by subtracting the previous equations:
\begin{equation}\label{Hfunction}
    H(r)\equiv \frac{4m'(r)\ell}{r^2}-\frac{2m''(r) \ell}{r}=-2E^2(r)L_F
\end{equation}
which will be used in the following sections to determine the validity of the solutions in the Maxwell regime.

Analogously, for the BTZ black hole, the field equations become:
\begin{align}
     \Lambda(r)-\Lambda_0 -\frac{M'(r)}{2r}&=\frac{1}{2}L(F)+ 2E^2(r) L_F  \\
   \Lambda(r) -\Lambda_0-\frac{M''(r)}{2}&=\frac{1}{2}L(F) 
\end{align}
and the geometric function:
\begin{equation}
  H(r) \equiv  \frac{M'(r)}{2r}-\frac{M''(r)}{2}
\end{equation}

Furthermore, we can solve the generalized Maxwell equation \eqref{mawxellgen}:
\begin{equation}\label{current}
    J^t = \frac{1}{\sqrt{-g}}\partial_{\alpha}\left(\sqrt{-g} L_F F^{\alpha t}\right) \implies J^t = -\frac{1}{r^2}\frac{d}{dr}\left[r^2 L_F E(r)\right],
\end{equation}
and from Eq. \eqref{lambda} we obtain:
\begin{equation}
    \partial_r \Lambda(r) = \frac{E(r)}{r^2}\frac{d}{dr}\left[r^2 L_F E(r)\right],
\end{equation}
demonstrating that, in this framework, the presence of the matter-induced cosmological contribution $\Lambda(r)$ can act as an effective electromagnetic source.

\section{Maxwell Fields as a source}\label{sec4}

In this section, we consider the linear Maxwell limit, $L(F)=-F$ and $L_F=-1$, within the non-conservative sector introduced above. Here, the Maxwell limit refers to the linear electromagnetic Lagrangian and does not imply a source-free field, since the effective current $J^\mu$ is in general nonvanishing. We will show that regular black string and $(2+1)$ black hole solutions can be supported by Maxwell electrodynamics in this sector.

We will do this by taking the linear limit of the Lagrangian: $L(F)=-F$ and $L_F=-1$. Thus:
\begin{equation}\label{Tmaxwell}
 T_{\mu \nu} = -g_{\mu\nu}\frac{F}{2} +2F_{\mu \alpha} F_{\nu}^{\ \alpha}.
\end{equation}
 Therefore, the field equations for black strings become:
\begin{align}
   \Lambda(r) -\Lambda_0 -\frac{4 m'(r) \ell}{r^2} &= -E^2(r) , \\ 
  \Lambda(r)  - \Lambda_0 - \frac{2 m''(r) \ell}{r} &= E^2(r) .
\end{align}

Since the electromagnetic energy-momentum tensor in the Maxwell limit is traceless, we have
\begin{equation}
 \Lambda(r)=  \Lambda_0 +\frac{ m''(r)\ell}{r}+\frac{2  m'(r) \ell}{r^2} ,
\end{equation}
thus, in the Maxwell limit, from the mass function, we can determine the  matter-induced cosmological contribution.

In three dimensions, the Maxwell energy-momentum tensor is no longer traceless. As a result, we have

\begin{equation}
\Lambda(r)=\Lambda_0+\frac{M''(r)}{6}+\frac{M'(r)}{3r}-\frac{E^2(r)}{3}.
\end{equation}

Combining this relation with the Einstein field equations,

\begin{align}
\Lambda(r)-\Lambda_0-\frac{M'(r)}{2r}&=-E^2(r),\\
\Lambda(r)-\Lambda_0-\frac{M''(r)}{2}&=E^2(r),
\end{align}
one obtains explicit expressions for both the electric field and $\Lambda(r)$ solely in terms of the mass function \(M(r)\):

\begin{align}
E^2(r)&=\frac{M'(r)}{4r}-\frac{M''(r)}{4},\\
\Lambda(r)&=\Lambda_0+\frac{M''(r)}{4}+\frac{M'(r)}{4r}.
\end{align}

The previously defined function $H(r)$, in the Maxwell limit, assumes the form
\begin{equation}
    H(r)=2E^2(r).
\end{equation}
Since the electric field must be real, Maxwell electrodynamics only supports the solutions if $H(r) \geq0$. 

Now, in the following subsections, we will show regular solutions supported by Maxwell electrodynamics. Results for spherical symmetry have already been obtained in Ref.\cite{3137491}.

\subsection{Bardeen-type Black String} \label{A}
The Bardeen spacetime is well known in the literature, and it was the first work to describe a regular black hole \cite{Bardeen1968}. In cylindrical symmetry, the electrically charged source that generates this spacetime was obtained in \cite{alencar2026blackstring}, and the Lagrangian is given by
\begin{equation}
        L(r)=12 \mu \ell ^3 \left(\frac{1}{q^2 \ell ^2+r^2}\right)^{7/2} \left(3 q^2 r^2-2 q^4 \ell ^2\right).
\end{equation}
Consequently, the mass function is given by
\begin{equation}
    m(r)=\frac{\mu r^3}{\left(q^2 \ell ^2+r^2\right)^{3/2}} 
\end{equation}
where $\mu$ is the vacuum linear mass density of the black string and $q$ is related to the electric charge. Using the definition of $H(r)$ in Eq. \eqref{Hfunction}, we have
\begin{equation}
    H(r)=\frac{30 \mu  q^2 r^2 \ell^3 }{\left(q^2 \ell ^2+r^2\right)^{7/2}},
\end{equation}
since $H(r)>0 \ \forall \ r$, the electric field $E(r)$ remains real, and the solution can be supported by Maxwell electrodynamics. Having $H(r)$, we can write the electric field as
\begin{equation}
    E^2(r)=\frac{15 \mu q^2 r^2 \ell^3}
    {\left(q^2\ell^2+r^2\right)^{7/2}}.
\end{equation}
In the limit $r\to0$, we find $E(r)\sim r$, whereas for
$r\to\infty$, $E(r)\sim r^{-5/2}$. Thus, the electric field
vanishes linearly at the origin, indicating that the electromagnetic
field is regular and strongly suppressed in the core. This behavior
prevents the Coulomb-like divergence at the center and contributes to
the regularity of the spacetime. At large distances, the field
decays as $r^{-5/2}$, ensuring that the electromagnetic contribution
becomes negligible asymptotically. Finally, calculating the matter-induced cosmological contribution, we obtain
\begin{equation}
\delta\Lambda(r)=-\frac{3 \mu  \ell ^3 \left(q^2 r^2-4 q^4 \ell ^2\right)}{\left(q^2 \ell ^2+r^2\right)^{7/2}}
\end{equation}
    In the absence of charges $(q\to0)$, $\delta\Lambda(r)=0 \implies \Lambda(r)=\Lambda_0$, which is the cosmological constant already used in the literature for black strings. We can see that $\delta\Lambda(r)$ remains regular at the origin,
\begin{equation}
\delta\Lambda(0)=\frac{12\mu}{q^3\ell^2}.
\end{equation}
Taking $\Lambda_0=-3/\ell^2$, this becomes
\begin{equation}
\Lambda(0)=\frac{3}{\ell^2}\left(\frac{4\mu}{q^3}-1\right).
\end{equation}
Therefore, the sign of $\Lambda(0)$ depends on the relative values of $\mu$ and $q$. For $\mu>q^3/4$, the core has $\Lambda(0)>0$ and is effectively de Sitter, while for $\mu<q^3/4$, $\Lambda(0)<0$ and the core is effectively anti-de Sitter. The limiting case $\mu=q^3/4$ corresponds to a locally Minkowski core. As $r\to\infty$, the spacetime recovers the standard AdS cosmological constant for black strings,
\begin{equation}
\delta\Lambda(r\to\infty)=0, \implies\Lambda(r\to\infty)=\Lambda_0.
\end{equation}

\subsection{Hayward-type Black String}\label{B}
The Hayward spacetime is a well-known spherically symmetric regular black hole model. It asymptotically recovers the Schwarzschild metric as $r \to \infty$, while exhibiting a de Sitter-like core at the origin ($r \to 0$) to avoid the central singularity \cite{Hayward:2005gi}. Originally, Hayward did not specify the physical matter or field source required to support such a geometry. Later, Fan and Wang \cite{Fan:2016hvf} successfully derived the corresponding source Lagrangian by employing a general model of nonlinear electrodynamics:
\begin{equation} \label{hayward}
    L(F) = -\frac{3 (\alpha F)^{3/2}}{\alpha \left[1+(\alpha F)^{3/4}\right]^2}.
\end{equation}

In cylindrical  symmetry, we use the solution obtained in Ref. \cite{alencar2026blackstring}, whose mass function is
\begin{equation}
    m(r)=\frac{\mu r^3}{r^3 + \lambda^3} 
\end{equation}
where $\lambda$ is related to the electric charge. Calculating the geometric function $H(r)$:
\begin{equation}
    H(r)=\frac{36 \lambda ^3 \mu  r^3 \ell }{\left(\lambda ^3+r^3\right)^3}
\end{equation}
and we can see that $H(r)$ remains positive for all $r$, confirming the validity of the electric source in the Maxwell regime. Now, we can write the electric field:
\begin{equation}
    E^2(r)=\frac{18\lambda^3\mu r^3\ell}
    {\left(\lambda^3+r^3\right)^3},
\end{equation}
which near the origin behaves as $E(r)\sim r^{3/2}$, while asymptotically
$E(r)\sim r^{-3}$. Thus, the electric field vanishes more rapidly near
the origin than in the Bardeen-type configuration, for which $E(r)\sim r$.
This indicates a stronger suppression of the electromagnetic field in
the Hayward-type core, contributing to a softer electromagnetic
contribution near the center.

Finally, calculating $\delta\Lambda(r)$, we find that
\begin{equation}
    \delta\Lambda(r) =  - \frac{6\lambda^3 \mu \ell \left(r^3 - 2\lambda^3\right)}
{\left(\lambda^3 + r^3\right)^3}.
\end{equation}
At the origin we have
\begin{equation}
\delta\Lambda(0)=\frac{12\mu\ell}{\lambda^3},
\end{equation}
which remains regular and $\Lambda(r)$ assumes the role of an effective cosmological constant, $\Lambda(r)=\Lambda_0+{12\mu\ell}/{\lambda^3}$. For $\Lambda_0=-3/\ell^2$, its sign is determined by the relative values of $\mu$ and $\lambda$, such that
\begin{equation}
\Lambda(0)
\begin{cases}
>0, & \mu>\dfrac{\lambda^3}{4\ell^3},\\[6pt]
=0, & \mu=\dfrac{\lambda^3}{4\ell^3},\\[6pt]
<0, & \mu<\dfrac{\lambda^3}{4\ell^3}.
\end{cases}
\end{equation}
Thus, the core is effectively de Sitter, Minkowski, or anti-de Sitter, respectively. In the asymptotic limit $r\to\infty$, we recover the standard cosmological constant,
\begin{equation}
\Lambda(r\to\infty)=\Lambda_0.
\end{equation}

\subsection{Regular Charged  $(2+1)$ Black Hole}\label{C}
Another example of a regular solution is the one obtained by Cataldo and Garcia for the BTZ-like black hole \cite{Cataldo:2000ns}, which is constructed using nonlinear electrodynamics. The mass function is given by
\begin{equation}
  M(r)=m+q^2\ln(r^2 + a^2)
\end{equation}
where $a$ is an adjustable parameter that regularizes the solution.

 Using the unimodular gravity (UG) procedure, we obtain

\begin{equation}
H(r)=\frac{2 q^2 r^2}{\left(a^2+r^2\right)^2},
\end{equation}

\begin{equation}
E^2(r)=\frac{q^2 r^2}{\left(a^2+r^2\right)^2}.
\end{equation}

It is straightforward to verify that $H(r)\geq0$ for all $r$, with $H(0)=0$. Therefore, the solution is well defined within Maxwell electrodynamics throughout the entire spacetime.

Near the origin, the electric field behaves as
\begin{equation}
    E(r)\approx\frac{q\,r}{a^2}+\mathcal{O}\left(r^3\right).
\end{equation}

At large distances, the electric field asymptotically approaches
\begin{equation}
    E(r)\approx\frac{q}{r},
\end{equation}
thereby recovering the standard Maxwell behavior in three-dimensional spacetime.

Finally, we obtain the matter-induced cosmological contribution:
\begin{equation}
   \delta \Lambda(r)=\frac{a^2 q^2}{\left(a^2+r^2\right)^2}
\end{equation} 

In the absence of charges, $\delta\Lambda(r)=0$, and the AdS cosmological constant is recovered, $\Lambda(r)=\Lambda_0$. Near the origin, we obtain a shifted cosmological constant, as in the previous cases, $\Lambda(r)=q^2/a^2+\Lambda_0$. Asymptotically, the AdS cosmological constant is recovered once again.

\subsection{Generating Maxwell-Supported Regular Solutions} \label{generatingregularsolutions}

The previous examples reveal a useful property of the UG--Maxwell system. Since the geometric function $H(r)$ is linear in the mass profile, different Maxwell-supported regular solutions can be combined to generate further solutions. As shown below, this property holds for both the black-string and $(2+1)$-dimensional sectors. It is not generically inherited by nonlinear electrodynamics, where the nonlinear dependence of $L_F$ on the electromagnetic invariant prevents the corresponding combination of the electromagnetic fields.

\subsubsection{Black-string case}

For the black string, the geometric function defined in Eq.~(28) is
\begin{equation}
H[m](r) = \frac{4 m'(r)\ell}{r^2} - \frac{2 m''(r)\ell}{r}.
\end{equation}
This is a linear differential functional of $m(r)$. Therefore, given a set of mass profiles $m_i(r)$ and positive coefficients $c_i$, the combined profile
\begin{equation}
m(r)=\sum_{i=1}^{N}c_i\, m_i(r)
\end{equation}
satisfies
\begin{equation}
H[m](r)=\sum_{i=1}^{N}c_i\, H[m_i](r).
\end{equation}
In the Maxwell limit, $H[m_i](r)=2E_i^2(r)$ for each seed solution, so that
\begin{equation}
H[m](r)=\sum_{i=1}^{N}c_i\,2E_i^2(r)=2\sum_{i=1}^{N}c_i\,E_i^2(r)\geq0,
\end{equation}
provided $c_i\geq0$ and $E_i^2(r)\geq0$ for every $i$. The combined geometry is therefore also supported by a real Maxwell field, with
\begin{equation}
E^2(r)=\sum_{i=1}^{N}c_i\,E_i^2(r).
\end{equation}

The regularity and asymptotic behavior are preserved provided that the seed solutions have the corresponding properties. If, near the origin,
\begin{equation}
m_i(r)=\alpha_i\, r^3+\mathcal{O}\!\left(r^{3+\delta_i}\right),
\qquad \delta_i>0,
\end{equation}
then
\begin{equation}
m(r)=\left(\sum_{i=1}^{N}c_i\alpha_i\right)r^3+\cdots,
\end{equation}
and the regular behavior at the origin is maintained. Similarly, if
\begin{equation}
m_i(r)\longrightarrow\mu_i \qquad (r\rightarrow\infty),
\end{equation}
then
\begin{equation}
m(r)\longrightarrow\mu_{\rm eff},
\qquad
\mu_{\rm eff}=\sum_{i=1}^{N}c_i\mu_i,
\end{equation}
and the Lemos asymptotic behavior is recovered with an effective linear mass density $\mu_{\rm eff}$. By the same linearity, the matter-induced cosmological contribution of the combined solution is
\begin{equation}
\Lambda(r)-\Lambda_0=\sum_{i=1}^{N}c_i\left[\Lambda_i(r)-\Lambda_0\right].
\end{equation}

As a direct example, consider the Bardeen-type and Hayward-type black strings discussed above. Their mass profiles can be combined as
\begin{equation}
m(r)=
c_B\,\frac{\mu_B r^3}{(q^2\ell^2+r^2)^{3/2}}
+
c_H\,\frac{\mu_H r^3}{r^3+\lambda^3},
\qquad
c_B,c_H\geq0.
\end{equation}
Using the corresponding expressions for $H(r)$, we find
\begin{equation}
H(r)=
c_B\,\frac{30\mu_Bq^2r^2\ell^{3}}{(q^2\ell^2+r^2)^{7/2}}
+
c_H\,\frac{36\lambda^3\mu_Hr^3\ell}{(\lambda^3+r^3)^3}.
\end{equation}
Both contributions are non-negative for positive parameters. Hence,
\begin{equation}
H(r)\geq0,
\end{equation}
and the combined mass profile defines another regular black string supported by Maxwell electrodynamics. The corresponding electric field is
\begin{equation}
E^2(r)=
c_B\,\frac{15\mu_Bq^2r^2\ell^{3}}{(q^2\ell^2+r^2)^{7/2}}
+
c_H\,\frac{18\lambda^3\mu_Hr^3\ell}{(\lambda^3+r^3)^3}.
\end{equation}
Thus, the Bardeen-type and Hayward-type solutions are not restricted to isolated examples, but can be used as seed geometries to generate a larger class of regular Maxwell-supported black strings.

\subsubsection{$(2+1)$ case}

The same construction can be applied to the $(2+1)$-dimensional sector. In this case, the geometric function is
\begin{equation}
H[M](r)=\frac{M'(r)}{2r}-\frac{M''(r)}{2}.
\end{equation}
Again, $H$ is linear in the mass profile. Thus, for
\begin{equation}
M(r)=\sum_{i=1}^{N}c_i\,M_i(r),
\end{equation}
we have
\begin{equation}
H[M](r)=\sum_{i=1}^{N}c_i\,H[M_i](r).
\end{equation}
In the Maxwell limit, $H[M_i](r)=2E_i^2(r)$, and therefore
\begin{equation}
E^2(r)=\sum_{i=1}^{N}c_i\,E_i^2(r).
\end{equation}
Consequently, if $c_i\geq0$ and each seed solution satisfies $H[M_i](r)\geq0$, the combined profile also satisfies
\begin{equation}
H[M](r)\geq0.
\end{equation}
The matter-induced cosmological contribution is again given by the linear superposition
\begin{equation}
\Lambda(r)-\Lambda_0=\sum_{i=1}^{N}c_i\left[\Lambda_i(r)-\Lambda_0\right].
\end{equation}
Therefore, the generating procedure remains applicable to $(2+1)$-dimensional geometries, provided that the seed profiles satisfy the regularity and asymptotic conditions required by the corresponding geometry.
\subsubsection{Comparison with nonlinear electrodynamics}

The above property relies not only on the linear dependence of $H$ on the mass profile, but also on the linear Maxwell relation between $H$ and $E^2$. This becomes clear when one considers a general nonlinear electromagnetic theory.

For a generic Lagrangian $L(F)$, Eq.~(28) gives the exact (non-Maxwell) relation
\begin{equation}
H(r)=-2E^2(r)\,L_F(F),
\qquad F=-2E^2(r),
\end{equation}
which we write as $H(r)=\mathcal{P}\!\left(E^2(r)\right)$, with
\begin{equation}
\mathcal{P}(x)\equiv-2x\,L_F(-2x).
\end{equation}
Because $L_F$ depends on $F$, and hence on $x=E^2$, the map $\mathcal{P}$ is generally nonlinear. For a combined mass profile $m(r)=\sum_i c_i m_i(r)$, linearity of $H[m]$ in $m(r)$ still gives
\begin{equation}
H[m](r)=\sum_i c_i\, H[m_i](r)=\sum_i c_i\,\mathcal{P}\!\left(E_i^2(r)\right),
\end{equation}
but this is \emph{not} the same as evaluating $\mathcal{P}$ on the combined field. For a generic nonlinear theory,
\begin{equation}
\mathcal{P}\!\left(\sum_i c_iE_i^2(r)\right)
\neq
\sum_i c_i\,\mathcal{P}\!\left(E_i^2(r)\right),
\end{equation}
and therefore the electric field of the combined geometry is not, in general, given by
\begin{equation}
E^2(r)=\sum_i c_i\,E_i^2(r).
\end{equation}
Instead, it must be obtained by solving $\mathcal{P}(E^2(r)) = \sum_i c_i\, \mathcal P(E_i^2(r))$ for $E^2(r)$, i.e., by inverting the nonlinear constitutive relation for the combined profile. Thus, even when two geometries are individually supported by a given NED model, their linear combination is not automatically supported by the same nonlinear theory.

The Maxwell case is different because $L_F=-1$ is constant, so $\mathcal{P}(x)=2x$ is itself linear. The constitutive relation reduces to
\begin{equation}
H(r)=2E^2(r),
\end{equation}
and the linearity of $H$ in the mass profile is directly inherited by $E^2$. Positive linear combinations of Maxwell-supported seed profiles therefore remain Maxwell-supported. This gives a simple way of generating classes of regular solutions in the UG--Maxwell framework from the individual geometries considered above, and highlights that the mechanism is specific to the linear (Maxwell) source rather than a generic feature of electromagnetically-supported regular geometries.

\section{Physical Properties}
\label{sec:physical_properties}

Having established the construction of regular Maxwell-supported geometries in the previous sections, we now turn to their physical properties. We therefore focus here on the features that are specific to the present UG construction and, in particular, on the new class of solutions generated in Sec.~\eqref{generatingregularsolutions}.

We first analyze the energy conditions associated with the matter-induced cosmological contribution $\delta\Lambda(r)$, which isolates the sector specific to the non-conservative UG framework. We then verify the regularity of the curvature invariants and study the thermodynamic properties of the generated black-string solution. This provides a physical characterization complementary to the solution-generating procedure developed above.

\subsection{Energy conditions of the matter-induced sector}
\label{sec:energy_conditions}

We now examine the energy conditions associated specifically with the matter-induced cosmological contribution,
\begin{equation}
    \delta\Lambda(r)=\Lambda(r)-\Lambda_0,
\end{equation}
which is the new ingredient of the present UG construction. The energy conditions of the individual regular geometries used as seeds have already been investigated in their original contexts, so we restrict the discussion here to the contribution generated by the non-conservative sector. 
The analysis of the weak energy condition for the $(2+1)$-dimensional case has already been carried out in \cite{Cataldo:2000ns}.

Separating the constant cosmological term, the field equations can be written as
\begin{equation}
    G_{\mu\nu}+\Lambda_0 g_{\mu\nu}
    =
    T_{\mu\nu}+T_{\mu\nu}^{(\mathrm{UG})},
    \end{equation}
where
\begin{equation}
    T_{\mu\nu}^{(\mathrm{UG})}
    =
    -\delta\Lambda(r)g_{\mu\nu}.
\end{equation}
Therefore, in an orthonormal frame,
\begin{equation}
    \rho_{\mathrm{UG}}=\delta\Lambda(r),
    \qquad
    p_{r}^{\mathrm{UG}}
    =
    p_{t}^{\mathrm{UG}}
    =
    -\delta\Lambda(r).
\end{equation}
It follows immediately that
\begin{equation}
    \rho_{\mathrm{UG}}+p_i^{\mathrm{UG}}=0,
\end{equation}
and hence the null energy condition is identically saturated by this sector.

The weak energy condition additionally requires
\begin{equation}
    \rho_{\mathrm{UG}}\geq0,
\end{equation}
and therefore reduces to
\begin{equation}
    \delta\Lambda(r)\geq0.
    \label{WECUG}
\end{equation}
Thus, the sign of the matter-induced cosmological contribution completely determines the WEC of the UG sector.

For the black-string geometry, the result obtained in the previous section gives
\begin{equation}
    \delta\Lambda(r)
    =
    \ell\left[
    \frac{m''(r)}{r}
    +
    \frac{2m'(r)}{r^2}
    \right].
\end{equation}
Since $\ell>0$, Eq.~\eqref{WECUG} is equivalent to
\begin{equation}
    r m''(r)+2m'(r)\geq0,
\end{equation}
or, more compactly,
\begin{equation}
    \left[r^2m'(r)\right]'\geq0.
    \label{WECmass}
\end{equation}
Equation~\eqref{WECmass} provides a general criterion for the matter-induced sector directly in terms of the mass profile. For the Bardeen case, we have
\begin{equation}
    \left[r^2m'(r)\right]'=
    \frac{3 \mu q^2\ell^2 r^3\left(4q^2\ell^2-r^2\right)}
    {\left(q^2\ell^2+r^2\right)^{7/2}}
    \geq 0
    \iff
    0\leq r\leq 2\ell q,
\end{equation}
which shows that the WEC is satisfied within the interval $0\leq r\leq 2\ell q$. In the asymptotic regime, $\left[r^2m'(r)\right]'\to0$, thereby recovering the validity of the WEC.

For Hayward case, we have
\begin{equation}
    \left[r^2m'(r)\right]'
    =
    -\frac{6\lambda^3\mu r^3\left(r^3-2\lambda^3\right)}
    {\left(\lambda^3+r^3\right)^3}
    \geq 0
    \iff
    0\leq r\leq 2^{1/3}\lambda,
\end{equation}
which shows that the WEC is satisfied within the interval
$0\leq r\leq 2^{1/3}\lambda$. In the asymptotic regime,
$\left[r^2m'(r)\right]'\to0$, recovering the WEC asymptotically.

An immediate consequence is that this property is compatible with the generating procedure introduced in Sec.~\eqref{generatingregularsolutions}. For a positive linear combination
\begin{equation}
    m(r)=\sum_i c_i m_i(r),
    \qquad c_i\geq0,
\end{equation}
one has
\begin{equation}
    \left[r^2m'(r)\right]'
    =
    \sum_i c_i
    \left[r^2m_i'(r)\right]'.
\end{equation}
Therefore, if each seed satisfies the WEC condition \eqref{WECmass}, any positive linear combination of these seeds also satisfies it. This provides a direct physical criterion, in addition to regularity and Maxwell support, for the class of solutions generated in Sec.\eqref{generatingregularsolutions}

\subsection{Curvature regularity}
\label{sec:curvature_regular}

We next verify explicitly that the generating procedure introduced in Sec.~\eqref{generatingregularsolutions} preserves the regularity of the geometry. Since the individual seed solutions are already known to be regular, the relevant question here is whether their combination can introduce a curvature singularity.

For the black-string metric \eqref{metric}, with
\begin{equation}
f(r)=\frac{r^2}{\ell^2}-\frac{4m(r)\ell}{r},
\end{equation}
regularity at the origin requires the mass function to behave as
\begin{equation}
m(r)=k r^3+\mathcal{O}(r^4),
\qquad r\rightarrow0.
\label{regularmass}
\end{equation}
where $k$ is a finite constant. Consequently,
\begin{equation}
f(r)=
\left(\frac{1}{\ell^2}-4k\ell\right)r^2
+\mathcal{O}(r^3).
\label{regularf}
\end{equation}
Thus, near the core, the metric function exhibits the same quadratic behavior as that of a constant-curvature geometry.

For the metric above, the Ricci scalar is
\begin{equation}
R=
-f''(r)-\frac{4f'(r)}{r}
-\frac{2f(r)}{r^2},
\end{equation}
while the quadratic invariants can be written as
\begin{equation}
R_{\mu\nu}R^{\mu\nu}
=
\frac{1}{2}
\left(
f''+\frac{2f'}{r}
\right)^2
+
2\left(
\frac{f'}{r}+\frac{f}{r^2}
\right)^2,
\end{equation}
and
\begin{equation}
R_{\mu\nu\rho\sigma}R^{\mu\nu\rho\sigma}
=
f''(r)^2
+
4\frac{f'(r)^2}{r^2}
+
4\frac{f(r)^2}{r^4}.
\end{equation}
Using Eq.~\eqref{regularf}, all potentially divergent combinations,
\begin{equation}
f''(r),\qquad
\frac{f'(r)}{r},
\qquad
\frac{f(r)}{r^2},
\end{equation}
remain finite as $r\rightarrow0$. Therefore,
\begin{equation}
\lim_{r\to0}R<\infty,
\qquad
\lim_{r\to0}R_{\mu\nu}R^{\mu\nu}<\infty,
\qquad
\lim_{r\to0}
R_{\mu\nu\rho\sigma}R^{\mu\nu\rho\sigma}<\infty.
\end{equation}

More explicitly, defining
\begin{equation}
a=\frac{1}{\ell^2}-4C\ell ,
\end{equation}
so that $f(r)=ar^2+\mathcal{O}(r^3)$, one obtains
\begin{equation}
R(0)=-12a,
\end{equation}
\begin{equation}
R_{\mu\nu}R^{\mu\nu}\big|_{r=0}
=36a^2,
\end{equation}
and
\begin{equation}
R_{\mu\nu\rho\sigma}R^{\mu\nu\rho\sigma}\big|_{r=0}
=24a^2.
\end{equation}
The core is therefore a finite constant-curvature region.

This result is immediately applicable to the generated solutions of Sec.~\eqref{generatingregularsolutions}. If
\begin{equation}
m(r)=\sum_i c_i m_i(r),
\end{equation}
and every regular seed satisfies
\begin{equation}
m_i(r)=k_ir^3+\mathcal{O}(r^4),
\end{equation}
then
\begin{equation}
m(r)=
\left(\sum_i c_i k_i\right)r^3
+\mathcal{O}(r^4).
\end{equation}
Hence the generated mass profile automatically preserves the condition
$m(r)=\mathcal{O}(r^3)$ at the origin. In particular, the combined
Bardeen--Hayward solution constructed in Sec.~\eqref{generatingregularsolutions} remains regular and
possesses finite Ricci, Ricci-squared, and Kretschmann invariants at its
core.

Thus, the linear generating procedure preserves not only Maxwell support
but also the curvature regularity of the seed geometries.

\subsection{Thermodynamics of the generated black string}
\label{sec:thermodynamics}

We finally consider the thermodynamic properties of the black-string solution generated in Sec.~\eqref{generatingregularsolutions}. For convenience, the combined Bardeen--Hayward mass profile can be parametrized as
\begin{equation}
m(r)=\mu r^3 {\cal A}(r),
\end{equation}
where
\begin{equation}
{\cal A}(r)=
\frac{\alpha}{(q^2\ell^2+r^2)^{3/2}}
+
\frac{1-\alpha}{r^3+\lambda^3},
\qquad
0\leq\alpha\leq1.
\label{Afunction}
\end{equation}
With this normalization, $m(r)\rightarrow\mu$ as $r\rightarrow\infty$, so that $\mu$ retains the interpretation of the asymptotic linear mass density. The limits $\alpha=1$ and $\alpha=0$ recover the Bardeen-type and Hayward-type seeds, respectively.

The metric function is
\begin{equation}
f(r)=\frac{r^2}{\ell^2}
-4\mu\ell r^2{\cal A}(r).
\end{equation}
For a nonvanishing horizon radius $r_h$, the condition $f(r_h)=0$ gives
\begin{equation}
\mu=
\frac{1}{4\ell^3{\cal A}(r_h)}.
\label{muhorizon}
\end{equation}
Thus, for fixed $q$, $\lambda$, and $\alpha$, the mass density can be expressed directly in terms of the horizon radius.

The Hawking temperature is
\begin{equation}
T_H=\frac{f'(r_h)}{4\pi}.
\end{equation}
Using Eq.~\eqref{muhorizon}, this reduces to
\begin{equation}
T_H=
-\frac{r_h^2}{4\pi\ell^2}
\frac{{\cal A}'(r_h)}{{\cal A}(r_h)}.
\label{THgeneral}
\end{equation}
For the combined solution,
\begin{equation}
{\cal A}'(r)=
-\frac{3\alpha r}{(q^2\ell^2+r^2)^{5/2}}
-\frac{3(1-\alpha)r^2}{(r^3+\lambda^3)^2},
\end{equation}
and therefore
\begin{equation}
T_H=
\frac{3}{4\pi\ell^2{\cal A}(r_h)}
\left[
\frac{\alpha r_h^3}
{(q^2\ell^2+r_h^2)^{5/2}}
+
\frac{(1-\alpha)r_h^4}
{(r_h^3+\lambda^3)^2}
\right].
\label{THcombined}
\end{equation}
For positive parameters and $r_h>0$, the temperature is non-negative.

The Bekenstein--Hawking entropy is proportional to the horizon area. Since the black string is extended along the $z$ direction, it is convenient to introduce the entropy density $s=S/L_z$. The horizon area per unit coordinate length is
\begin{equation}
\frac{A_h}{L_z}
=
\frac{2\pi r_h^2}{\ell},
\end{equation}
and hence
\begin{equation}
s=
\frac{\pi r_h^2}{2G\ell}.
\label{entropyDensity}
\end{equation}

The heat capacity density at fixed $q$, $\lambda$, and $\alpha$ can then be defined by
\begin{equation}
c_{q,\lambda,\alpha}
=
T_H
\left(
\frac{\partial s}{\partial T_H}
\right)_{q,\lambda,\alpha}.
\end{equation}
Using $r_h$ as the thermodynamic parameter, this becomes
\begin{equation}
c_{q,\lambda,\alpha}
=
T_H
\frac{ds/dr_h}{dT_H/dr_h}.
\end{equation}
From Eq.~\eqref{THgeneral}, one obtains
\begin{equation}
c_{q,\lambda,\alpha}
=
\frac{
\pi r_h^3{\cal A}(r_h){\cal A}'(r_h)
}{
G\ell
\left\{
{\cal A}(r_h)
\left[
2r_h{\cal A}'(r_h)
+r_h^2{\cal A}''(r_h)
\right]
-r_h^2[{\cal A}'(r_h)]^2
\right\}
}.
\label{heatcapacitycombined}
\end{equation}
Here,
\begin{equation}
{\cal A}''(r)=
-\frac{3\alpha(q^2\ell^2-4r^2)}
{(q^2\ell^2+r^2)^{7/2}}
+
\frac{6(1-\alpha)r(2r^3-\lambda^3)}
{(r^3+\lambda^3)^3}.
\end{equation}

Equation~\eqref{heatcapacitycombined} gives the heat capacity explicitly for the generated family. Its sign determines the local thermodynamic stability: $c_{q,\lambda,\alpha}>0$ corresponds to a locally stable branch, whereas $c_{q,\lambda,\alpha}<0$ corresponds to an unstable one. Possible divergences occur when
\begin{equation}
{\cal A}(r_h)
\left[
2r_h{\cal A}'(r_h)
+r_h^2{\cal A}''(r_h)
\right]
-r_h^2[{\cal A}'(r_h)]^2=0,
\end{equation}
which is equivalent to an extremum of the Hawking temperature and separates different thermodynamic branches.

The parameter $\alpha$ continuously interpolates between the Bardeen-type and Hayward-type limits. Therefore, Eqs.~\eqref{muhorizon}, \eqref{THcombined}, and \eqref{heatcapacitycombined} provide the horizon structure and local thermodynamic behavior of the complete family generated by their combination.

\section{Conclusions} \label{sec5}
In this work, we show that regular black string solutions can be supported by Maxwell electrodynamics within the framework of unimodular gravity. We have seen that the condition $\det(g_{\mu\nu})=const.$ reduces the set of spacetime symmetries, allowing the conditions on the energy-momentum tensor to be relaxed, such that it is not necessarily conserved, implying now an integration function $\Lambda\to\Lambda(x)$. A recent result has shown that the condition on $\det(g_{\mu\nu})$ arises directly from the Einstein Equivalence Principle, being compatible with the framework of unimodular gravity. 
The non-conservative sector explored here is a property of the UG framework considered in this work and should not be understood as a consequence of the EEP. The connection with the EEP discussed in Ref.~[1] concerns specifically the fixed-volume condition underlying unimodular gravity.

We constructed the black string solution proposed by Lemos \cite{LEMOS} within the framework of unimodular gravity, as well as the BTZ black hole \cite{Banados:1992wn}, and appropriately defined the integration constant so that it corresponds to the AdS cosmological constant $\Lambda_0$. Using nonlinear electrodynamics, we found that the integration function acts as an effective electromagnetic source \cite{3137491}. We also found that it is not possible (in our configuration) to obtain purely electric sources, since this would imply $\Lambda=\Lambda_0$.

From the definition of the geometric function $H(r)$, we determine the validity of the solutions in the Maxwell regime, which requires $H(r)\geq 0$. We find that this condition is satisfied for all the solutions considered, as well as for their corresponding linear combinations. This latter result is particularly interesting, since in the context of nonlinear electrodynamics (NED), a linear combination of mass functions is not necessarily supported by the same NED theory, whereas in the Maxwell case considered here, the corresponding linear combination remains consistently Maxwell-supported. This allows for a more general analysis of the physical properties of the resulting family of solutions, including the energy conditions, curvature invariants, and thermodynamic properties.

A natural extension of the present analysis is to investigate the dynamics of this class of fields, with particular emphasis on the stability of the resulting solutions under perturbations and on the behavior of the associated electromagnetic and gravitational fields. Such investigations may provide further insights into the dynamical properties and physical viability of the regular black-hole and black-string solutions obtained within this framework.

\acknowledgments{The author G.A. would like to thank Conselho Nacional de Desenvolvimento Cient\'{i}fico e Tecnol\'{o}gico (CNPq) and Fundação Cearense de Apoio ao Desenvolvimento Científico e Tecnológico (FUNCAP) through PRONEM PNE0112- 00085.01.00/16, for the partial financial support. V.H.U.B. is supported by Coordena\c c\~{a}o de Aperfei\c coamento de Pessoal de N\'{i}vel Superior - Brasil (CAPES).
}
\bibliographystyle{apsrev4-1}
\bibliography{referencias}
\end{document}